\documentclass[article,twocolumn,showpacs,preprintnumbers,amsmath,amssymb]{revtex4}
\usepackage[dvips]{color}
\usepackage{graphicx}% Include figure files
\usepackage{dcolumn}% Align table columns on decimal point
\usepackage{bm}% bold math
\usepackage{epsfig}

\usepackage{color}

 \newcommand{{\vp}}{{\vec p}}
\newcommand{{\vq}}{{\vec q}} 
 \newcommand{\beq}{\begin{equation}}
  \newcommand{\eeq}[1]{\label{#1} \end{equation}}

\newcommand{\lton}{\mathrel{\lower.9ex \hbox{$\stackrel{\displaystyle
        <}{\sim}$}}} \newcommand{\gton}{\mathrel{\lower.9ex
    \hbox{$\stackrel{\displaystyle >}{\sim}$}}}
\newcommand{\ee}{\end{equation}} \newcommand{\bea}{\begin{eqnarray}}
\newcommand{\eea}{\end{eqnarray}}
\newcommand{\beqar}{\begin{eqnarray}}
  \newcommand{\eeqar}[1]{\label{#1}\end{eqnarray}}

\begin{document}

\title{A possible determination of the quark radiation length 
in cold nuclear  matter}

\author{R. B. Neufeld$^1$}
\email{neufeld@lanl.gov}

\author{Ivan Vitev$^1$}
\email{ivitev@lanl.gov}

\author{Ben-Wei Zhang$^{2,1}$}
\email{bwzhang@iopp.ccnu.edu.cn}

\affiliation{ $^1$ Los Alamos National Laboratory, Theoretical
Division, Los Alamos, NM 87545, USA } %

\affiliation{ $^2$ Key Laboratory of Quark $\&$ Lepton Physics (Huazhong Normal
University), Ministry of Education, China } %

\vspace*{1cm}

\begin{abstract}

We calculate the differential Drell-Yan production cross section in proton-nucleus collisions
by including both next-to-leading order perturbative effects and effects of the nuclear
medium. We demonstrate that dilepton production in fixed target experiments is an excellent
tool to study initial-state parton energy loss in large nuclei and to accurately determine
the stopping power of cold nuclear matter. We provide theoretical predictions for the attenuation
of the Drell-Yan cross section at large values of Feynman $x_F$ and show that for low proton beam
energies experimental measurements at Fermilab's E906 can clearly distinguish between nuclear
shadowing and energy loss effects. If confirmed by data, our results may help determine the quark
radiation length in cold nuclear matter $X_0 \sim 10^{-13}$~m.

\end{abstract}


\maketitle

%%%%%%%%%%%%%%%%%%%%%%%%%%%%%%%%%%%%%%%%%%%%%%%%%%%%%%%%%%%%%%%%%%%%%%%%%%

\section{Introduction}

The energy loss of a charged particle as it traverses dense matter is a fundamental probe of the
matter properties. Accurate theoretical calculations and experimental measurements of this quantity
became one of the great early successes of the classical and quantum theories of electromagnetic
interactions~\cite{Yao:2006px}.  The stopping power, $dE / dx$, in
the limit of large energies is dominated by bremsstrahlung processes and is related to the
radiation length, $X_0$, of charged particles in matter as follows:
\begin{equation}
-dE/dx = E/X_0 \;.
\label{eloss}
\end{equation}
In Eq.~(\ref{eloss}) $X_0$ is approximately independent of the incident particle momentum.
Precise knowledge of the stopping power and radiation length of materials is widely used today
in X-ray tomography, muon and proton radiography,
and high energy nuclear and particle physics detector development and instrumentation.

The fundamental constituents of nuclei, quarks and gluons, predominantly interact via their
color charge. The forces between them are described from first principles by the theory of
strong interactions, Quantum Chromodynamics (QCD), and are much larger in magnitude than
the electromagnetic force. In the last decade, advances in high energy many-body QCD
have enabled exploration of parton energy
loss~\cite{Baier:1996kr,Gyulassy:2000fs,Wang:2001ifa,Arnold:2002ja} in
a novel hot and dense state of nuclear matter -  the quark-gluon plasma (QGP). The predicted
suppression of energetic particle production in nucleus-nucleus (A+A) collisions,
dubbed jet quenching, is now definitively established~\cite{Adler:2003qi,Adams:2003kv}.

\begin{figure}[!t]
\includegraphics[width = 0.85\linewidth]{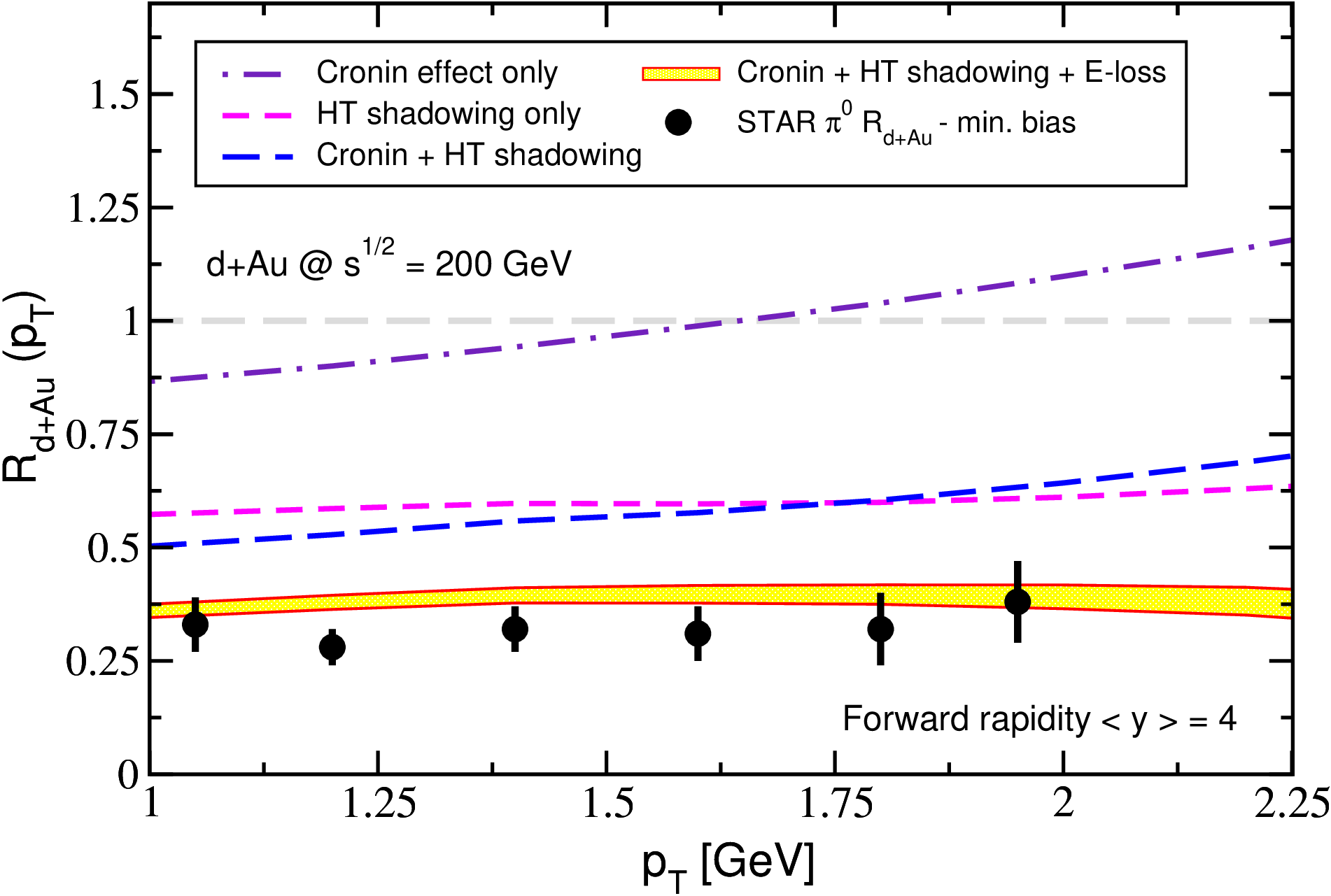}
\caption{Neutral pion suppression in minimum bias d+Au collisions at $y=4$ at RHIC.
Theoretical calculations that include known nuclear matter effects are shown. A complete
simulation gives a good description of the experimental data.}
\label{STARdA}
\end{figure}

Alongside the excitement of this new discovery comes the realization that one of the biggest gaps
in our current knowledge of nuclear reactions in extremis is the stopping power of cold nuclei for
color-charged particles. In d+Au collisions at the Relativistic Heavy Ion Collider (RHIC), where the QGP 
is not formed, a similarly large suppression of particle production at large forward rapidity is
observed~\cite{Arsene:2004ux,Adams:2006uz}. STAR experimental data on $\pi^0$ attenuation
at $y=4$, for example,  can be fitted with disparate  models that emphasize either
large shadowing or large nuclear stopping.  A more realistic calculation that includes the 
Cronin effect~\cite{Vitev:2003xu},
high-twist shadowing~\cite{Qiu:2004da}, and initial-state energy loss~\cite{Vitev:2007ve} -
all independently constrained in different processes and center of mass energies -
is presented in Figure~\ref{STARdA}. The incorporation of these effects in the perturbative QCD
calculation is briefly summarized in Ref.~\cite{Sharma:2009hn}. Our new results show that 
both coherent and inelastic scattering on the nucleus have comparable impact on the observed cross
section attenuation in this kinematic range. This finding stresses once again the need to identify
and reliably evaluate experimental observables that are clean and sensitive signatures
of jet energy loss in cold nuclear matter.

From a phenomenological point of view, the significance of cold nuclear matter energy loss is
well-established~\cite{Vogt:1999dw,Gavin:1991qk,Johnson:2000ph,Kopeliovich:2005ym}.
Theoretically, until recently, only final-state inelastic interactions in large nuclei
relevant to semi-inclusive deeply inelastic scattering have been considered~\cite{Wang:2002ri}.
A new comparative study of initial-state and final-state energy loss~\cite{Vitev:2007ve} suggests
that  in the limit of large parton energies these exhibit  strikingly different
path length  and parent parton energy  dependencies:
\begin{eqnarray}
\label{IS}
\Delta E^{\rm rad.}_{\rm initial-state} &\sim& \kappa_{LPM} \, C_R \alpha_s E  \frac{L}{\lambda_g}  \;,  \\
\Delta E^{\rm rad.}_{\rm final-state} &\sim& C_R \alpha_s  \frac{ \xi^2 L^2 }{\lambda_g} \; .
\label{FS}
\end{eqnarray}
Here, $C_R$ is the quadratic Casimir in the fundamental and adjoint representations for
quarks and gluons, respectively, $\lambda_g$ is the gluon mean free path of
${\cal O}(1\ {\rm fm})$ and  $\alpha_s$
is the strong coupling constant. In this paper we refer to the soft interactions with typical
transverse momentum transfers squared 
$\xi^2 \sim 0.1$~GeV$^2$ for quarks prior to the large $Q^2$ scattering as initial-state. 
Similarly, the soft interactions after the hard scattering are described  as final-state.  
We emphasize that such separation is only possible if $\xi^2 A^{1/3} \ll Q^2$. We will only be
interested in lepton pair production of invariant mass squared $M^2=Q^2 \geq 10$~GeV$^2$, which 
is compatible with this constraint. Finally, we point out that the $\xi^2$ dependence in 
Eq.~(\ref{IS}) is implicit through $\kappa_{LPM}$, the overall suppression factor relative to
the incoherent Bertsh-Gunion bremsstrahlung.

The differential medium-induced bremsstrahlung spectrum can be expressed as a solution of an
inhomogeneous recurrence relation with suitably chosen boundary conditions~\cite{Vitev:2007ve}.
These boundary conditions differ for initial-state and final-state energy loss and the results will be
correspondingly different. Final-state interactions and Eq.~(\ref{FS}) have been investigated 
in detail, for example see~\cite{Baier:1996kr,Gyulassy:2000fs,Wang:2001ifa,Arnold:2002ja}. Let us 
now focus on Eq.~(\ref{IS}). Our starting point is the integral form for the double 
differential medium-induced gluon bremsstrahlung spectrum~\cite{Vitev:2007ve}:
\begin{eqnarray}     
&& \!\!\!\!\! \!\! k^+ \frac{dN^g(IS)}{dk^+ d^2 {\bf k}}  =  \frac{C_R \alpha_s}{\pi^2}  
\int  d^2 {\bf q} \;  
\frac{\xi_{\rm eff}^2}{\pi ({\bf q}^2+\xi^2)^2} 
\left[   \frac{L}{\lambda_g}   \frac{{\bf q}^2}{{\bf k}^2 
({\bf k}-{\bf q})^2}   \right.
\nonumber \\
&& \hspace*{2.2cm} \left. -  2    \frac{{\bf q}^2 - 
{\bf k} \cdot {\bf q} }{{\bf k}^2 ({\bf k}-{\bf q})^2}    
  \frac{k^+}{ {\bf k}^2 \lambda_g}  
\sin \left( \frac{  {\bf k}^2  L}{k^+} \right)   \right]  \;, 
\label{HBG1}
\end{eqnarray}     
In Eq.~(\ref{HBG1}) ${\bf k}$ is the transverse momentum of the gluon relative to the direction of
the parent parton, $k^+$ is its large lightcone momentum and ${\bf q}$ is the momentum transfer from
the nuclear medium. The formation time of the gluon $\tau_f = k^+/{\bf k}^2$ in comparison to 
the size of the medium $L$ determines the degree of the destructive interference between the 
Bertsch-Gunion radiation and the
radiation from the hard scattering. Let us focus on $k^+ \sim E^+$  and recognize that when 
$\tau_f \ll L$  and ${\bf k}$ varies, the phase factor $\sin(L/\tau_f)$ oscillates    
rapidly and averages to zero. One is left with the first term in the integrand of Eq.~(\ref{HBG1}),
which is the incoherent medium-induced bremsstrahlung: 
\begin{eqnarray}     
&&  \!\!\!\!\!\!\! \Delta E^{\rm rad.}_{\rm initial-state}|_{ \frac{E^+}{{\bf k}^2} \ll L} 
=  \frac{C_R \alpha_s}{\pi^2}\frac{L}{\lambda_g} E \, 
\bigg\{ \int  d^2 {\bf k} \;   \int  d^2 {\bf q} \; \nonumber \\    
&& \hspace*{2.3cm} \times  \frac{\xi_{\rm eff}^2}{\pi ({\bf q}^2+\xi^2)^2}  
  \frac{ {\bf q}^2 }{{\bf k}^2 ({\bf k}-{\bf q})^2  }  \bigg \}. 
\label{approxBG}
\end{eqnarray} 
In the opposite limit $ \tau_f \gg L$  for $k^+ \sim E^+$ we can expand 
the sine function to lowest order and obtain: 
\begin{eqnarray}     
&&  \!\!\!\!\!\!\! \Delta E^{\rm rad.}_{\rm initial-state}|_{ \frac{E^+}{{\bf k}^2} \gg L}  
= \frac{C_R \alpha_s}{\pi^2}\frac{L}{\lambda_g} E \, 
\bigg\{ \int  d^2 {\bf k} \;   \int  d^2 {\bf q} \;\nonumber \\     
&& \hspace*{2.3cm} \frac{\xi_{\rm eff}^2}{\pi ({\bf q}^2+\xi^2)^2}  
 \, \left[ \frac{1}{({\bf k}-{\bf q})^2  } - \frac{1}{{\bf k}^2}    \right]  \bigg \}. 
\label{approxHBG}
\end{eqnarray} 
Note that the overall multiplicative coefficients $\{\cdots \}$ in Eqs.~(\ref{approxBG}) 
and~(\ref{approxHBG}) have to be evaluated numerically with the relevant kinematic cuts 
specified in Ref.~\cite{Vitev:2007ve}. In the coherent regime the coefficient also reflects the 
destructive interference effect between the bremsstrahlung associated with the soft scattering
and the bremsstrahlung associated with the large $Q^2$ process and can be numerically small. 
As the energy of the parent parton in the rest frame of the large nucleus grows, the
approximation for $\Delta E^{\rm rad.}_{\rm initial-state}$ given by Eq.~(\ref{approxHBG}) becomes
more relevant. This is the basis for the advocated  energy and path length dependence in
Eq.~(\ref{IS}).

Of course, there are always parts of the emitted gluon phase space $(k^+,{\bf k})$ that are 
not compatible with simple approximations. For this reason, we first 
evaluate the fully differential bremsstrahlung spectrum numerically from Eq.~(\ref{HBG1}), as 
described  in~\cite{Vitev:2007ve}. From Eq.~(\ref{eloss}) in the small energy loss limit we can 
then quote a radiation length:
\begin{equation} 
X_0 = L E \left[ \int dk^+  \int d^2{\bf k}  \,   
k^+ \frac{dN^g(E, L)}{dk^+ d^2 {\bf k}} \right]^{-1}  \, .
\label{radlength}
\end{equation}

To summarize, for final-state interactions, the destructive Landau-Pomeranchuk-Migdal
(LPM) interference leads to a change in the functional form of radiative energy loss. 
Eq.~(\ref{FS}) does not allow for a natural definition of a radiation length and implies 
that the experimentally  observable effects are limited to relatively small quark and gluon 
energies. In contrast, even if the LPM suppression
factor $\kappa_{LPM} \sim 1/10$ in Eq.~(\ref{IS}), $\Delta E^{\rm rad.}_{\rm initial-state} $ retains
some of the characteristics of incoherent bremsstrahlung, see Eqs.~(\ref{HBG1}), 
(\ref{approxBG}) and~(\ref{approxHBG}). 
For this reason, initial-state  energy loss can
also significantly affect experimental observables in heavy ion collider experiments of much
higher $\sqrt{s_{NN}}$~\cite{Sharma:2009hn,Vitev:2009rd,Vitev:2008vk}. Furthermore, 
Eq.~(\ref{IS}) implies that the stopping power of cold nuclear matter for partons
prior to a hard $Q^2 \gg \Lambda_{QCD}^2$ scattering can be characterized by a radiation 
length $X_0$ defined in Eq.~(\ref{radlength}). One can see parametrically 
from Eq. (\ref{IS}) 
that $X_0$ is expected to be  of  ${\cal O}(10\ {\rm fm} - 100 \ {\rm fm})$ - the 
shortest radiation length in nature,  ten orders of magnitude smaller than the radiation 
length of high-$Z$ materials, such as tungsten, for electrons.

The Drell-Yan process in heavy ion collisions~\footnote{In this paper we will use the 
term heavy ion collisions to 
describe both p+A and A+A hadronic reactions at relativistic energies}
- $q + \bar{q} \rightarrow \gamma^* \rightarrow l^+ + l^-$ at lowest order (LO) -
is an ideal probe of initial-state effects. The final-state particles do not
interact strongly with the nuclear medium, providing a relatively clean experimental signature.
Still, definitive separation of leading-twist shadowing
effects~\cite{Arneodo:1992wf,Hirai:2004wq,Armesto:2006ph} and  parton energy
loss~\cite{Vitev:2007ve,Arleo:2002ph,Garvey:2002sn,Duan:2003yc,Wang:2007gt}
has so far proven challenging~\cite{Vasilev:1999fa,Johnson:2001xfa}. In no small part this difficulty
arises from the fact that the very same Drell-Yan data in proton-nucleus (p+A) reactions is used 
to constrain
the shadowing parameterizations~\cite{Eskola:1998df}. Going to higher center of mass energies,
such as what is available at RHIC and the Large Hadron Collider (LHC), will only amplify this 
unfortunate ambiguity.
In this paper we demonstrate how dilepton production in conjunction with a suitably low
proton beam energy $E_{\rm beam} = 120$~GeV, corresponding to $\sqrt{s_{NN}}=15$~GeV, at Fermilab's 
experiment
E906 can disentangle initial-state energy loss and nuclear shadowing effects and pinpoint the 
stopping power of
cold nuclear matter for quarks and gluons with $\sim 20\%$ accuracy.

Our work is organized as follows. In Section~\ref{sec:p+p} we investigate lepton pair
production in hadron-hadron reactions at next-to-leading order (NLO)
in collinear factorized perturbative QCD and validate the theoretical simulation tools 
against experimental
data.  In Section~\ref{sec:CNM} we discuss the cold nuclear matter (CNM) effects:
initial-state energy loss and nuclear shadowing.
A complete calculation of the dilepton  production rate in p+A  collisions at NLO that includes
CNM effects is given in Section~\ref{sec:p+A}. We compare results to existing data on
the attenuation of the Drell-Yan cross section  in reactions with nuclei and present
predictions for this suppression versus the cold nuclear matter radiation length
$X_0$ for the upcoming E906 measurements. A summary and conclusions are presented in
Section~\ref{conclude}.

\section{The Drell-Yan process in p+p collisions }
\label{sec:p+p}

In this paper we will be interested in the differential Drell-Yan production cross
sections $d\sigma^{DY}/dx_F d Q^2$ and $d\sigma^{DY}/dy dQ^2$ where  $Q^2 = q^2= (p_{l^+}+p_{l^-})^2$
is the invariant mass squared, $x_F = 2 q_L /\sqrt{s}$ is the Feynman $x_F$, and
$y=\frac{1}{2}\ln [(q_0+q_L)/(q_0-q_L)]$ is the rapidity of the lepton pair.
For illustration, we start the discussion with the LO cross section in the collinear factorization
approach~\cite{Collins:1989gx}:
\begin{eqnarray}\label{dylodx}
 \frac{d\sigma^{DY}}{dx_FdQ^2} = \sum_{q,{\bar q}} \frac{4\pi e_q^2\alpha^2}
{9 Q^2 s (x_1 + x_2)} f_{q(\bar q)}(x_1,\mu)f_{\bar q(q)}(x_2,\mu) \; .
\end{eqnarray}
In Eq.~(\ref{dylodx}) $\alpha \approx 1/137$, $e_q$ is the quark fractional electric charge and
$f_i(x_i,\mu)$ are the parton distribution functions. The incident parton  lightcone momentum
fractions $x_1,x_2$ obey the relations:
\begin{equation}
x_1 x_2 = Q^2/s\;, \quad x_F = x_1 - x_2 \,.
\end{equation}

\begin{figure}
\includegraphics[width = 0.95\linewidth]{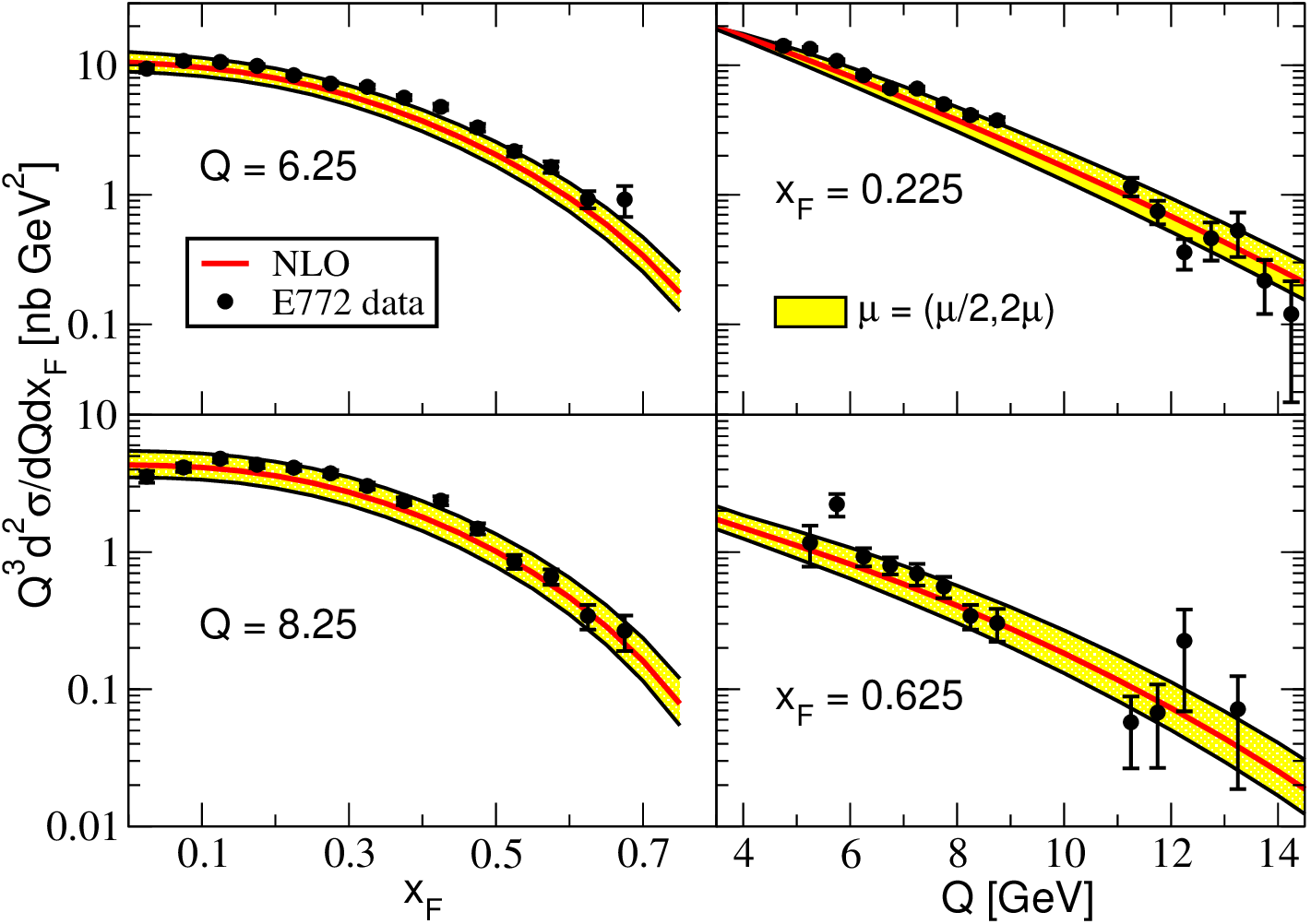}
\caption{Next-to-leading order  calculations for the muon pair production cross section in 
p+D collisions at
$\sqrt{s_{NN}} = 38.8$~GeV is compared to Fermilab E772 experimental data. The yellow bands
represent the theoretical uncertainty due to choice of factorization and renormalization scales.
We have only included isospin effects for the deuterium target.}
\label{DYpp}
\end{figure}

It is well-known that the perturbative next-to-leading order corrections to the
Drell-Yan process can be quite significant~\cite{Altarelli:1979ub,Kubar:1980zv,Stirling:1993gc}.
The NLO calculation also adds a Compton scattering contribution to the annihilation graphs.
Schematically, the one-loop cross section can be written as:
\begin{equation}
\frac{d\sigma^{DY}_{NLO}}{dQ^2dx_F} =  \frac{d\sigma^{DY}_A}{dQ^2dx_F} 
+ \frac{d\sigma^{DY}_C}{dQ^2dx_F} \;.
\label{DYNLO}
\end{equation}
The importance of going to ${\cal O}(\alpha^2 \alpha_s)$ is two-fold. First, it removes 
phenomenological
$K_{NLO}$ factors and allows a more reliable evaluation of the perturbative uncertainty through the
variation of the renormalization and factorization scales $\mu_R=\mu_f=\mu$. Second, it facilitates
a more accurate incorporation of dynamical nuclear effects that scale with the quadratic Casimir, which
plays the role of the average squared color charge for quarks and gluons. In this work we implement the
NLO Drell-Yan cross section calculation following Ref.~\cite{Kubar:1980zv}.

An alternative approach to lepton pair production is based on the dipole 
model~\cite{Raufeisen:2002zp,Betemps:2003je}. In this model the Drell-Yan process
is viewed as induced virtual photon bremsstrahlung in the target rest frame and 
nuclear effects are parameterized in the dipole cross section. In this
paper we adhere to the traditional QCD factorization approach where higher-order
corrections can be systematically evaluated order-by-order in perturbation theory.   
Furthermore, nuclear effects in perturbative QCD are incorporated at the cross section level
as opposed to the amplitude level and are thus dependent on the average path length of
partons through dense matter~\footnote{For example, in a picture of interfereing virtual 
photon production
amplitudes from distinct nucleon-nucleon scatterings the path length will be the average
of the two path lengths. }.

To test the baseline perturbative calculation for the Drell-Yan production cross section we
compare our simulation results to Fermilab E772 data~\cite{McGaughey:1994dx} from a fixed
deuterium target experiment with an incident proton beam energy $E_{\rm beam} = 800$~GeV in
Figure~\ref{DYpp}. We take into account isospin effects as described in~\cite{Vitev:2005he}.
The left panels of Figure~\ref{DYpp} show results for two different values of the lepton pair
mass $Q$. The right panels show results for two different values of $x_F$. The yellow 
band represents
the perturbative uncertainty due to the variation of the renormalization and factorization
scales $Q/2 \leq \mu \leq 2 Q$. We have checked that shadowing or energy loss effects 
for the deuterium target have less than a few \%
effect on the calculated cross section and are not shown in Figure~\ref{DYpp} for simplicity.

\section{Cold nuclear matter  effects}
\label{sec:CNM}

As seen in Figure~\ref{STARdA}, cold nuclear matter effects can significantly alter the
production cross section for energetic/massive final states. Universal initial-state  
leading-twist shadowing effects are incorporated in nuclear parton distributions (nPFDs).
Application for different observables or different $\sqrt{s_{NN}}$ is achieved by
multiplying the standard parton distribution functions (PDFs) with a nPDF correction
factor~\cite{Vitev:2005he}. Alternatively, one can attempt to calculate CNM effects associated 
with the elastic, inelastic and coherent initial-state and final-state parton interactions 
in large nuclei from first principles~\cite{Vitev:2006bi}. This latter approach aims to elucidate 
the physics that underlays the observed nuclear modification. To date, there is no global nPDF
analysis that attempts to separate the process-dependent and process-independent effects, largely 
due to the limited knowledge of the latter. Better theoretical and experimental control on 
cold nuclear matter energy loss is one step to help rectify this deficiency.

The process-dependent nuclear effects in question are enumerated below.
First comes the Cronin effect that is often modeled through initial-state transverse momentum 
broadening~\cite{Accardi:2002ik}
and affects the $q_T$ distributions of energetic particles. In this manuscript we consider
$q_T$-integrated cross sections, see for example Eqs.~(\ref{dylodx}) and (\ref{DYNLO}),
and do not include the Cronin effect. Next comes dynamical shadowing that arises 
from the coherent final-state interactions of the recoil parton in the
nuclear target~\cite{Qiu:2004da}. For the Drell-Yan process at LO this effect vanishes 
since there is no parton in the final state.
At NLO there is indeed a parton in the final state, however, for the dilepton
masses of interest $Q = 4 - 10$~GeV the power suppressed high-twist shadowing effects,
$\propto (A^{1/3} \xi^2)/Q^2 $ where $\xi^2 \sim 0.1$~GeV$^2$, are expected to be negligible. 
This leaves the stopping power of large nuclei for incoming quarks and gluons as the most 
significant dynamical nuclear effect for Drell-Yan production. We discuss the
cold nuclear matter energy loss below since it is central to  our paper.

We evaluate initial-state energy loss numerically for minimum bias p+A reactions with
deuterium ($^2$D), beryllium ($^9$Be), carbon ($^{12}$C), aluminum ($^{27}$Al), iron ($^{56}$Fe),
and tungsten ($^{184}$W) targets following Ref.~\cite{Vitev:2007ve}. We use $\lambda_g=1$~fm
for the gluon mean free path and vary the momentum transfer per interaction with the medium from
$\xi = 0.175$~GeV to $\xi = 0.5$~GeV~\footnote{The momentum transfer between
the jet and the medium is often denoted by $\mu$. We use $\xi$ in this paper to indicate 
that the default value corresponds to the same dimensional scale used in dynamical shadowing
calculations.}. Comparing our results to Eq.~(\ref{radlength}) we find that
in the limit of small energy loss $\Delta E^{\rm rad}_{\rm initial-state} \ll E$ or, equivalently,
$L \ll X_0$, the radiation length of 100~GeV quarks in cold nuclear matter is between
30~fm and 160~fm for the parameters discussed above. We also calculate the probability
distribution $P_{q,g}(\epsilon)$ for quarks and gluons to lose a fraction of their energy
$\epsilon = \sum_i \omega_i / E$ due to multiple gluon emission in the Poisson
approximation~\cite{Vitev:2005he,Baier:2001yt}:   
\begin{eqnarray}
\label{psn}
 \int_0^1P_{q,g}(\epsilon) \, d\epsilon &=& 1\;,  \\ 
\quad \int_0^1P_{q,g}(\epsilon) \epsilon \, d\epsilon &=& 
\frac{\Delta E^{\rm rad.}_{q,g\, {\rm initial-state}}}{E} \; .
\label{poisson}
\end{eqnarray}

The calculation and/or implementation of initial-state or final-state state energy loss processes 
focuses on real medium-induced gluon emission. Let us consider for definitiveness the energy loss 
of the first incoming parton (characterized by lightcone momentum $p^+_1$ and fraction $x_1$).
The differential distribution of emitted gluons itself is evaluated by separating it from the partonic 
scattering cross section. For a general hard process:  
\begin{eqnarray}
\frac{d \sigma}{dPS} &=& \int dx_1 dx_2 \,  f(x_1) f(x_2)  \int d \epsilon  \, 
\frac{ \langle |M[p^+_1(1-\epsilon),p^+_2]  |^2 \rangle }{2 x_1(1-\epsilon) x_2 s} \nonumber \\
&& \times 
 (2\pi)^4 \delta^4(p_i   - p_f)  \frac{dN^g(\epsilon)}{d\epsilon}    \, .
\label{radscat}
\end{eqnarray}
In Eq.~(\ref{radscat}) $\epsilon = k^+/p_1^+$ and parton flavor indices and the sum over
flavors are not shown explicitly.
We emphasize that the energy loss results only depend on the details of the final state through
the value of $p_1^+$ and in the $\delta$-function $ p_i = p_1(1-\epsilon) + p_2$. We have
also integrated over the gluon's transverse momentum ${\bf k}$, noting that the direction
of the incoming parton does not change on average. In the soft gluon approximation, which 
we use in this study, medium-induced bremsstrahlung can be identified even at the amplitude 
level~\cite{Gyulassy:2000fs,Vitev:2007ve}. Proper factorization of the differential energy
loss distribution in the $\epsilon \rightarrow 1$ limit can only be shown in QCD at the cross
section level~\cite{Ovanesyan:2011xy}. This distribution does include the interference of 
the bremsstrahlung from the soft in-medium scattering with the bremsstrahlung from the 
large $Q^2$ process.

For a single ($n_g=1$) emitted gluon $dN^g(\epsilon)/d\epsilon = P(\epsilon)$ is also the 
probability distribution for fractional energy loss $\epsilon$. In general, $n_g \neq 1$
and the probability distributions $P_{q,g}(\epsilon)$, see Eqs.~(\ref{psn}) and~(\ref{poisson}),
is constricted from $dN^g(\epsilon)/d\epsilon$ in the independent Poisson 
approximation~\cite{Vitev:2005he,Baier:2001yt} to gluon emission.       
Changing variables, $x_1 \rightarrow x_1 / (1-\epsilon)$, and the order of the 
integration we find:
\begin{eqnarray}
\frac{d \sigma}{dPS} &=& \int dx_1 dx_2  \left[ \int d \epsilon \, P(\epsilon)  
f\left(\frac{x_1}{1-\epsilon} \right)  \right]  f(x_2)
\nonumber \\
 && \times \frac{1}{2 x_1 x_2 s} \langle |M [p^+_1,p^+_2]|^2 \rangle
 (2\pi)^4 \delta^4(p_i-p_f) \, . \quad
\label{chvb}
\end{eqnarray}

From Eq.~(\ref{chvb}), initial-state energy loss is most conveniently implemented in
the perturbative calculations of the differential
Drell-Yan cross section Eqs.~(\ref{dylodx}) and~(\ref{DYNLO}) as follows:
\begin{eqnarray}
\label{ISloss}
f_{q,{\bar q}}(x,\mu) & \rightarrow  & \int_0^1 d\epsilon \, P_q(\epsilon) f_{q,{\bar q}}
\left(\frac{x}{1-\epsilon},\mu \right) \;,   \\
f_{g}(x,\mu) & \rightarrow & \int_0^1 d\epsilon \, P_g(\epsilon) f_{g}
\left(\frac{x}{1-\epsilon},\mu \right) \;.
\label{IS1}
\end{eqnarray}
Here,  $f(x,\mu)$ is the distribution function of the parton from the incident proton.
The physical meaning of  Eqs.~(\ref{ISloss}) and (\ref{IS1})  is that in the presence of nuclear
stopping larger quark and gluon energies in the nuclear wave function are probed by the
same lepton pair kinematics. These results imply that the attenuation of the Drell-Yan cross section 
depends not only on the magnitude of cold nuclear matter energy loss but also on the
steepness of the PDFs. Specifically, large suppression is expected at large values
of Feynman $x_F$ (or large rapidity) where $f^\prime(x,\mu) \ll 0$. We have denoted by $f^\prime(x,\mu)$ 
the derivative of the parton distribution function with respect to the momentum fraction $x$.
Finally, we note that accounting for fluctuations
in the energy loss via $P_{q,g}(\epsilon)$ is critical when jet and particle production
is near kinematic limits. We have checked that using the mean energy loss
$\Delta E^{\rm rad.}_{q,g\, {\rm initial-state}} / E $, instead of the full convolution 
over $P_{q,g}(\epsilon)$, overpredicts the suppression at large $x_F$ by more than a factor of two.

\section{Lepton pair  production in p+A collisions}
\label{sec:p+A}

\begin{figure}[!t]
\includegraphics[width = 0.9\linewidth]{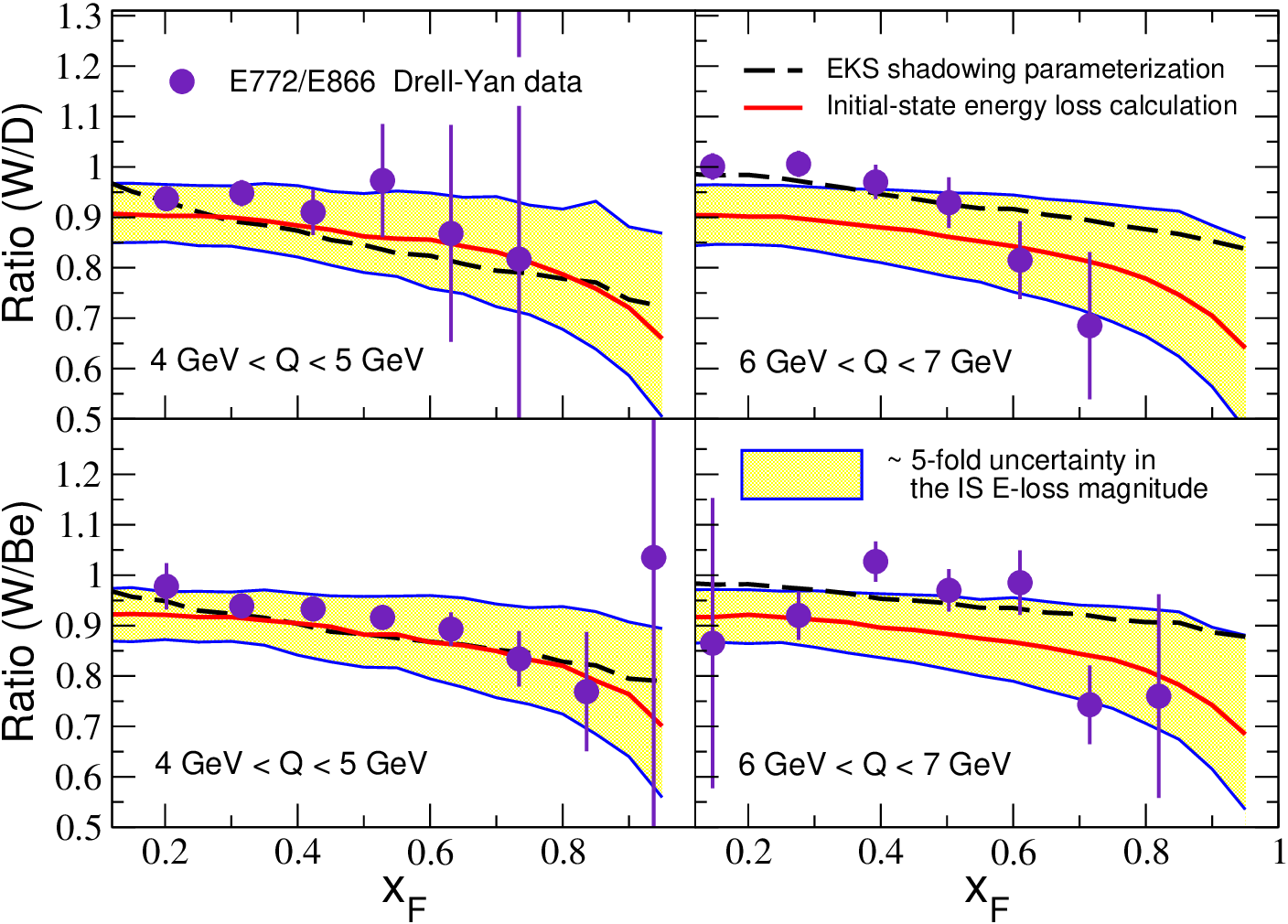}
\caption{Plot of the ratio of the dimuon cross section $\sigma(\text{p+W})/\sigma(\text{p+A})$,
where A = D,Be, from both experiment and theory for 800~GeV
protons colliding with a fixed nuclear target. The theory curves are the NLO Drell-Yan cross section
calculation, including shadowing effects and initial-state energy loss, respectively.
The yellow bands indicate a five-fold variation of the
mean quark medium-induced energy loss. }
\label{pA800}
\end{figure}

We now present results from the full calculation of Drell-Yan production at NLO that
include cold nuclear matter effects. We first consider existing measurements from the
Fermilab  E772/E886 experiment~\cite{Johnson:2000ph} which were performed using several nuclear targets
in the path of a proton beam of energy $E_{\rm lab} = 800$~GeV. Dilepton data and
theoretical simulations are presented in Figure~\ref{pA800} as the ratio of the differential cross sections
on two different targets in minimum bias p+A reactions scaled down to a binary
nucleon-nucleon interaction:
\begin{equation}
{\rm Ratio}(A/B) \equiv  R_{AB}^{DY}
= B \cdot \frac{d\sigma^{DY}_{pA}}{dx_F dQ^2} \Big/ A \cdot \frac{d\sigma^{DY}_{pB}}{dx_F dQ^2} \;.
\label{RAA}
\end{equation}
We considered two different invariant mass ranges
$4~{\rm GeV}< Q< 5~{\rm GeV}$ (left panels) and   $6~{\rm GeV}< Q< 7~{\rm GeV}$ (right panels).
We also included two different ratios: $R_{WD}^{DY}$ and $R_{WBe}^{DY}$ in the top and bottom
panels, respectively. The dashed lines represent a calculation of the cross section ratios
that incorporates a leading twist-shadowing parameterization~\cite{Eskola:1998df}. The solid lines
show our simulation with initial-state energy loss~\cite{Vitev:2007ve} that is compatible with
the theoretical description of RHIC d+A and A+A data~\cite{Sharma:2009hn,Vitev:2009rd,Vitev:2008vk}.

It should be noted that while both calculations describe the E772/E866 data with 
equal acumen, they emphasize completely different physics. The former simulation
assumes a modification of the nuclear wave-function and is driven by the decreasing
momentum fraction $x_2$ of the parton in the nucleus when Feynman $x_F $ grows. The
behavior of $R_{AB}^{DY}$ for the latter case is governed by the energy loss of the
incident parton that carries an increasing fraction $x_1$ of the momentum of the proton. The
current accuracy of the experimental data does not allow one to distinguish between
these two competing paradigms. Furthermore, it cannot precisely constrain the magnitude
of $\Delta E_{\rm initial-state}^{\rm rad.}$. The yellow band in Figure~\ref{pA800}
represents a variation of the typical momentum transfers between the incoming parton
and the medium $\xi$ in the range of 0.175~GeV to 0.5~GeV,
as described in Section~\ref{sec:CNM}. Such variation induces
an approximately five-fold uncertainty in the magnitude of the stopping power of large
nuclei for incident partons. While the extreme cases of very large or very small
energy loss are not favored by the data, a considerable uncertainty can still
exist for this physics scenario.

A unique opportunity to disentangle initial-state energy loss effects from
nuclear shadowing is presented by the approved and now under construction
Fermilab experiment E906~\cite{web,Reimer:2007iy}. It is a fixed target Drell-Yan
experiment with $E_{\rm beam} = 120$~GeV.  Its primary goal is to accurately
measure the large $x_F$ $\bar{d}/\bar{u}$ asymmetry and it can comfortably
cover the $0.4 \leq x_F \leq 0.9$  and  $M_{J/\psi} \leq Q \leq M_{\Upsilon}$
kinematic range. A secondary goal of E906 is to precisely determine cold nuclear
matter energy loss. Its low center of mass energy per nucleon pair $\sqrt{s_{NN}}=15$~GeV
allows a selection of the dilepton mass which practically eliminates shadowing
effects.  In contract, nuclear stopping is expected to be even more important at
these energies.

Next, let us review  again the prerequisites for the determination of the stopping power of
cold nuclear matter and its radiation length $X_0$. First, one needs to identify the
part of phase space where shadowing effects are minimal (we are interested specifically
in the Drell-Yan process). Second, experimental measurements must establish a statistically 
significant suppression of the dilepton production cross section in p+A reactions  
$R_{pA}^{DY}<1$ at large $x_F$. Third, one has to confirm an approximately linear dependence 
of $R_{pA}^{DY}$  on the nuclear size $A^{1/3}$. Note that the $A^{1/3}$ dependence of the  
$\Delta E^{\rm rad.}$  is correlated with a linear dependence on the parton energy. This will 
allow for a proper  extraction of $X_0$.

\begin{figure}
\includegraphics[width = 0.9\linewidth]{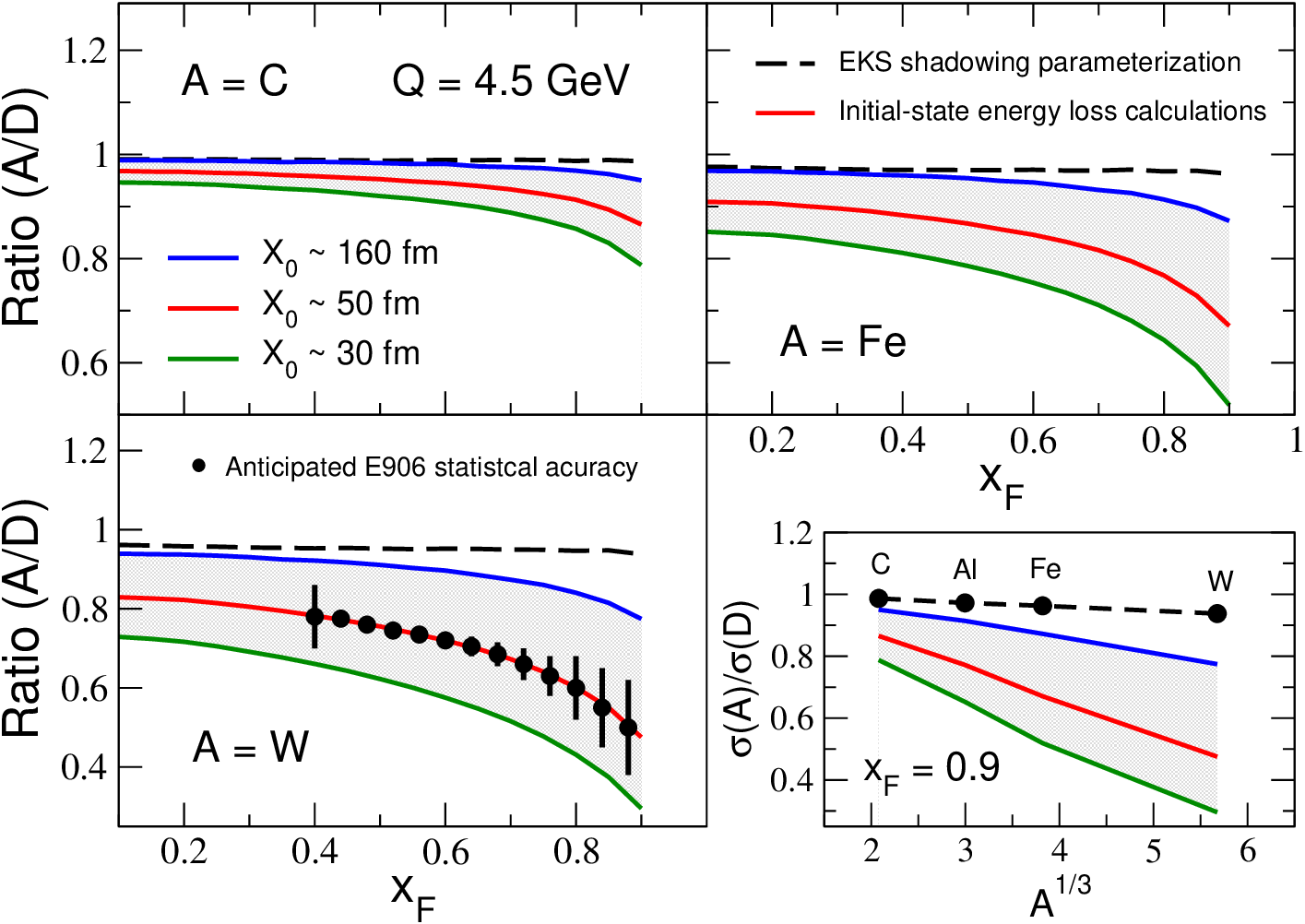}
\caption{Next-to-leading order theoretical predictions for the Drell-Yan dimuon cross section
attenuation $\sigma(\text{p+A})/\sigma(\text{p+D})$  at Fermilab's E906.
As with Figure~\ref{pA800}, the curves include shadowing
effects (dashed line) or initial-state energy loss (solid lines), respectively.
We have considered three different radiation lengths $X_0$ of quarks in large nuclei,
ranging from 30~fm to 160~fm.
The lower right panel highlights the  difference between the two physics
scenarios at large $x_F$ as a  function of $A^{1/3}$.}
\label{pA120}
\end{figure}

Our numerical results, relevant to the E906 program, are presented in Figure~\ref{pA120}.
We have chosen to consider $d\sigma^{DY}/dx_F dQ^2$ versus Feynman $x_F$ for $Q = 4.5$ GeV, 
and have carried out simulations for deuterium, carbon, iron, and tungsten targets.
In order to obtain the ratio defined in Eq.~(\ref{RAA}), we have normalized the nuclear 
cross sections to that of a D target which minimizes any trivial isospin dependence. In all cases,
the effects of shadowing, which are shown with dashed lines in Figure~\ref{pA120}, are
negligible. On the other hand, energy loss effects, represented by solid lines for three
different radiation lengths $X_0 = 30$~fm, 50~fm and 160~fm, can clearly be detected,
especially for large values of $x_F$. We have also included in Figure~\ref{pA120} simulated
experimental data which shows the anticipated E906 statistical precision for
$R_{WD}^{DY}$~\cite{Reimer:2007iy}. With careful selection of the heavy nuclear targets,
such as Fe and W, and the lepton pair kinematics, the two competing pictures for the physics
that underlays the suppression of dilepton production in p+A reactions can be definitively
confirmed or refuted. As seen in Figure~\ref{pA120}, the radiation length of cold nuclear
matter for quarks can likely be constrained with $\sim 20\%$ accuracy.

The bottom right panel in Figure~\ref{pA120} shows the attenuation of the Drell-Yan
cross section at a fixed value of $x_F=0.9$ versus the linear size $\sim A^{1/3}$
of the target nucleus. Our results reflect the fact that initial-state
energy loss of quarks and gluons prior to a large $Q^2$ scattering depends approximately
linearly on the path length through strongly interacting matter. This is in contrast
to final-state energy loss where quadratic path-length
dependence~\cite{Baier:1996kr,Gyulassy:2000fs,Wang:2001ifa} would suggest $A^{2/3}$
scaling of the attenuation of the lepton pair production rate. E906 measurements
will also help clarify this open question.

\section{Summary and conclusions}
\label{conclude}

One of the most pressing current questions in the theory and phenomenology of heavy ion
reactions at high energies is related to the stopping power of cold nuclear matter
for energetic quarks and gluons. Plentiful experimental evidence and theoretical justification
exist for the significance of initial-state energy loss
effects~\cite{Arsene:2004ux,Adams:2006uz,Gavin:1991qk,Johnson:2000ph,Kopeliovich:2005ym,Vitev:2007ve,Sharma:2009hn,Vitev:2009rd}.
It is, therefore, unfortunate that attempts to precisely determine their strength have
so far been inconclusive.
These earlier studies employed a functional form for the energy loss that might have been
better-suited to describe final-state scattering processes in the QGP, and yielded large
uncertainties for $-dE/dx$~\cite{Johnson:2000ph,Arleo:2002ph,Garvey:2002sn,Vasilev:1999fa}.
Little is known quantitatively for the interactions of partons and hadrons in cold
nuclear matter beyond the interaction lengths, or mean free paths, for selected
few species, such as protons or alpha particles. Only recently have developments in the theory of
quark and gluon propagation in cold nuclear matter suggested that
initial-state energy loss in QCD retains certain characteristics of induced electromagnetic
bremsstrahlung~\cite{Vitev:2007ve}. Thus, it can be described in terms of a
radiation length $X_0$. A real opportunity exists today to determine this shortest radiation length
in nature.

To this end, we have embarked on a program to developed the theoretical tools that can both
motivate and facilitate the upcoming E906 experimental p+A program. To ensure that the baseline
differential Drell-Yan cross sections in elementary nucleon-nucleon reactions are reliably estimated,
we opted for a next-to-leading order accuracy of the perturbative QCD calculation.
The new simulation tool was validated against Fermilab's E772 muon pair production
measurements in p+D reactions. Next, we implemented initial state-nuclear effects,
such as leading-twist shadowing and medium-induced bremsstrahlung in large nuclei. Care was taken to
incorporate the fluctuations in the magnitude of parton energy loss. We showed that neglect
of these fluctuations will result in a sever overestimate of the attenuation of the lepton
pair production rate. Results from our work have been used to evaluate
the dilepton background for the $\Upsilon$ measurements at RHIC via the $l^++l^-$
decay channel~\cite{private}.

For  nuclear targets with large mass number we first verified that existing Drell-Yan 
measurements cannot
distinguish between calculations that incorporate initial-state energy loss and leading-twist
shadowing parametrizations. While this observation has been made before, our work
is more general in that it goes beyond the fixed energy loss per unit length conjecture.
Our results also suggest that moving to the higher center-of-mass energies of RHIC and the LHC
will not help separate the two competing physics effects.
We showed that radiation lengths in cold nuclear matter for quarks $X_0$ of the order of 50~fm
to 100~fm are compatible with the $E_{\rm beam}=800$~GeV Fermilab E772/E866 data.
The same stopping power of large nuclei has worked well for the theoretical description
of hard probes production at RHIC.

To definitively confirm or refute initial-state energy loss as the principal
nuclear effect which leads to the suppression of the Drell-Yan cross section in
p+A reactions we presented predictions for the upcoming Fermilab experiment E906.
We exploited its low center-of-mass energy to identify a kinematic region where
shadowing effects vanish. We demonstrated that, with the anticipated statistical accuracy
of the data in the large $x_F$ region, radiation lengths as large as 200~fm
can be detected. If $X_0$ is of ${\cal O}(50~{\rm fm})$, there is a good chance that
it can be determined with $\sim20\%$ accuracy.

As a final note, we recognize that experimental measurements may differ from the
most sophisticated theoretical predictions. With the next-to-leading order simulation
tools at hand, upcoming E906 experimental data can be quickly
and reliably analyzed for parton energy loss effects. The theoretical progress described here,
combined with experimental advances in the use of dimuons, is expected to result in the
first unambiguous measurement of quark energy loss in nuclei~\cite{ldrd} and
provide the much-needed standard candle for gauging the nuclear response to strongly
interacting particles. \\

\begin{acknowledgments}
We thank M. Leitch and P. McGaughey for providing us with Fermilab E772/E866 experimental data.
This research is  supported by the US Department of Energy, Office of Science,
under Contract No. DE-AC52-06NA25396 and in part by the LDRD program at LANL,
by the Ministry of Education of China with the Program NCET-09-0411,
by National Natural Science Foundation of China with Project No. 11075062,
and CCNU with Project No. CCNU09A02001.
\end{acknowledgments}

\end{document}